\documentclass[runningheads]{svmult}

\usepackage{makeidx}   
\usepackage{graphicx}  
\usepackage{subeqnar}  
\usepackage{multicol}  
\usepackage{physprbb}  
\makeindex             

\begin{document}
\title*{Theory of electron spectroscopies}
\toctitle{Theory of electron spectroscopies}
%
\titlerunning{Theory of electron spectroscopies}
%
\author{Michael Potthoff}
\authorrunning{Michael Potthoff}
%
%
\institute{
Institut f\"ur Physik,
Humboldt-Universit\"at zu Berlin, 
Germany
}

\maketitle              

\begin{abstract}
The basic theory of photoemission, inverse photoemission, 
Auger-electron and appearance-potential spectroscopy is 
developed within a unified framework starting from Fermi's
golden rule.
The spin-resolved and temperature-dependent appearance-potential 
spectroscopy of band-ferromagnetic transition metals is studied in 
detail.
It is shown that the consideration of electron correlations and 
orbitally resolved transition-matrix elements is essential for a 
quantitative agreement with experiments for Ni.
\end{abstract}

\section{Basic electron spectroscopies}

\begin{figure}[t]
\centering \includegraphics[width=0.75\linewidth]{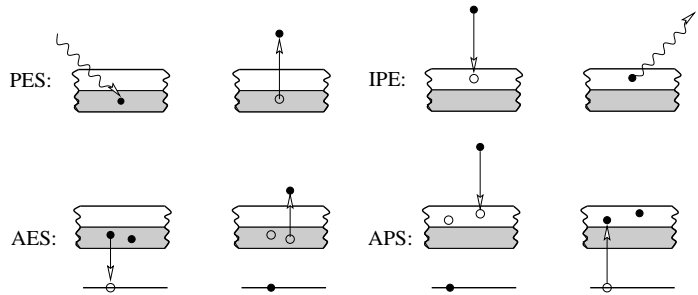}
\vspace{4mm}
\caption{
Schematic picture of the initial and the final states for different
electron spectroscopies.
PES: photoemission spectroscopy ($s=1$).
IPE: inverse photoemission ($s=-1$).
AES: Auger-electron spectroscopy ($s=2$).
APS: appearance-potential spectroscopy ($s=-2$).
The grey region is the occupied part of the valence band
}
\label{fig:specs}
\end{figure}

Electron spectroscopy \cite{Pen74,PH74,FFW78,Fug81,Ram91,Don94} 
is one of the fundamental experimental techniques to investigate the 
electronic structure of transition metals.
For a deeper understanding of the electronic properties, a meaningful 
interpretation of the measured spectra is necessary which is based on
a reliable theory of spectroscopies.
This is particularly important for the study of band ferromagnetism 
which is caused by a strong Coulomb interaction among the valence 
electrons.
Due to the presence of strong electron correlations, simple explanations 
of spectral features within an independent-particle model may fail.
Here we try to clarify what physical quantity is really 
measured and what ingredients are needed for a theoretical approach 
to find a satisfactory agreement with the experimental data.
We are concerned with the theory of four basic types of electron 
spectroscopy (see Fig.\ \ref{fig:specs}):
photoemission (PES), inverse photoemission (IPE), Auger-electron (AES) 
and appearance-potential spectroscopy (APS). 

The spin-, angle- and energy-resolved (ultraviolet) valence-band
photoemission (PES) more or less directly measures the occupied 
part of the band structure.
Present theories of PES are mostly based on the so-called one-step 
model of photoemission \cite{Pen74,Pen76,Bor85,Bra96} which treats the 
initial excitation step, 
the transport of the photoelectron to the surface and the scattering at 
the surface barrier as a single quantum-mechanically coherent process.
The one-step model is essentially based on the independent-particle 
approximation.
The valence electrons move independently in an effective potential 
as obtained from band-structure calculations within the local-density 
approximation (LDA) of density-functional theory (DFT) \cite{HK64,KS65}.
There are recent attempts \cite{PLNB97,MPN+99} for a reformulation 
of the one-step model
to include a non-local, complex and energy-dependent self-energy 
which accounts for electron correlations.

The one-step model also applies to inverse photoemission spectroscopy.
IPE is complementary to PES and yields information on the unoccupied 
bands above the Fermi energy \cite{Bra96}.
For example, IPE can be used to determine the dispersions of the 
Rydberg-like series of surface states in the $1/z$ image potential 
in front of the surface \cite{Don94}.

The theory of CVV Auger-electron spectroscopy (AES) is more complicated 
compared to PES/IPE since there are two valence electrons participating 
in the transition.
Due to core state involved additionally, AES is highly element specific.
This feature is frequently exploited for surface characterization.
Contrary to $\bf k$ resolved (inverse) photoemission, the Auger 
transition is more or less localized in real space.
High-resolution AES may thus yield valuable information on the local
valence density of states (DOS).
In the most simple theoretical approach suggested by Lander in 1953
\cite{Lan53}, the Auger spectrum is given by the self-convolution of 
the occupied part of the DOS.
Modern theories of AES also include the effect of transition-matrix 
elements \cite{HWMR88}.
Similar as the one-step model of PES and IPE, these approaches are based 
on the independent-electron approximation.

Finally, the appearance-potential spectroscopy (APS) is complementary 
to AES. 
Within Lander's independent-electron model \cite{Lan53} the AP line 
shape resulting from CVV transitions is given by the self-convolution 
of the unoccupied part of the DOS. 
Matrix elements are included in refined independent-particle theories
\cite{EP97}.
Its comparatively simple experimental setup and its surface sensitivity 
qualifies APS to study surface magnetism, for example.
For a ferromagnetic material, the spin dependence of the AP signal 
obtained by using a polarized electron beam gives an estimate of the 
surface magnetization as has been demonstrated for the transition 
metals Fe and Ni \cite{Kir84,EVDN93}.

Common to all four spectroscopies are a number of fundamental concepts and 
approximation schemes used in the theoretical description.
Therefore, it should be possible to develop the basic theory of PES, IPE,
AES and APS within a unified framework.
This is discussed in the following section \ref{sec:specs}.
As a result of this basic theory of electron spectroscopies it found that
in each case the intensity is essentially given in terms of a characteristic
Green function.

The actual evaluation of the Green function is more specific and depends on 
the respective spectroscopy, on the material and the physical question to 
be investigated. 
In section \ref{sec:ferro} we will therefore pick up an example and give a 
more detailed discussion of APS for band ferro-magnets.

A notable quality of APS is its direct sensitivity to electron correlations:
In the AP transition, two valence electrons are added to the system at a 
certain lattice site.
These final-state electrons mutually experience the strong intra-atomic
Coulomb interaction $U$.
As a consequence the two-particle (AP) excitation spectrum is expected to
show up a pronounced spectral-weight transfer or even a satellite with a 
characteristic energy of the order of $U$ \cite{Cin77,Saw77}.
This is an important difference between the one- (PES, IPE) and the two-particle 
spectroscopies (AES, APS).
To demonstrate this sensitivity to correlations in APS we consider the spin- 
and temperature-dependent AP spectra of ferromagnetic Nickel as a prototype 
of a strongly correlated and band-ferromagnetic material in section \ref{sec:ni}.
The discussion of the calculations and the comparison with experimental 
results will show what presently can be achieved in the theory of electron 
spectroscopies.
Some conclusions are given in section \ref{sec:con}.

\section{Theory}
\label{sec:specs}

The basic theory starts from the Hamiltonian $H$ which shall describe the 
electronic structure of the system within a certain energy range around the 
Fermi energy.
In addition, we consider a perturbation $V_s$ that mediates the respective 
transition. 
$V_s$ is assumed to be small and will be treated in lowest-order perturbation 
theory.
The index $s$ distinguishes the different spectroscopies 
(see Fig.\ \ref{fig:specs}).
$s=\pm 1,\pm 2$ stands for the difference between the number of valence 
electrons before and after the transition. 

In the photoemission spectroscopy (PES, $s=+1$) a valence electron is
excited into a high-energy scattering state by absorption of a photon. 
The photoelectron escapes into the vacuum and is captured by a detector
depending on its spin, energy and angles relative to the crystal surface.
Within the framework of second quantization, the perturbation can be
written as:
\begin{equation}
  V_{+1} = \sum_{\alpha\gamma} M_{\alpha\gamma} \, 
  a^\dagger_\alpha c_\gamma + \mbox{h.c.} \qquad \mbox{(PES)} \: .
\end{equation}
$c_\gamma$ annihilates a valence electron with quantum numbers $\gamma$,
and $a_\alpha^\dagger$ creates an electron in a high-energy scattering state 
$\alpha$.
The matrix element is calculated using the electric dipole approximation.
This is well justified in the visible and ultraviolet spectral range.
Neglecting the term quadratic in the field as usual and choosing the
Coulomb gauge, one obtains:
\begin{equation}
   M_{\alpha\gamma} = \langle \alpha | {\bf A}_0 {\bf p} | \gamma \rangle 
   \qquad (s=\pm 1) \: .
\end{equation}
Here $\bf p$ is the momentum operator and ${\bf A}_0$ is the spatially 
constant vector potential inside the crystal.
It can be determined from classical macroscopic dielectric theory.
Note that we use atomic units with $e=m_{\rm e}=\hbar=1$.

In an inverse-photoemission experiment (IPE, $s=-1$) an electron beam
characterized by quantum numbers $\alpha$ hits the crystal surface. 
An electron may be de-excited into an empty valence state $\gamma$ above 
the Fermi energy by emission of a photon.
The photon yield is measured as a function of $\alpha$ (the polarization, 
the energy and the angles of the incident electron beam). 
We have:
\begin{equation}
  V_{-1} = \sum_{\gamma\alpha} M_{\gamma\alpha}^\ast \, 
  c^\dagger_\gamma a_\alpha + \mbox{h.c.} \qquad \mbox{(IPE)} \: .
\end{equation}
Obviously, $V_{-1}=V_1$. 
Both, PES and IPE, are induced by the same electron-photon interaction. 

In the case of the two-particle ($s=\pm 2$) spectroscopies AES and APS, 
the transitions are radiationless.
Here the Coulomb interaction mediates between the initial and the final
state.
The initial state for AES ($s=+2$) is characterized by a hole in a 
core level (produced by a preceding core-electron excitation 
via x-ray absorption, for example). 
In the Auger transition the empty core state $\beta$ is filled by an
electron from a valence state $\gamma$. 
The energy difference is transferred to another valence electron in the 
state $\gamma'$ which is excited into a high-energy scattering state 
$\alpha$ and detected.
Thus,
\begin{eqnarray}
  V_{+2} = \sum_{\alpha \beta \gamma' \gamma}
  M_{\alpha\beta\gamma'\gamma} \, 
  a^\dagger_\alpha b^\dagger_\beta
  c_\gamma c_{\gamma'}  
  + \mbox{h.c.} \qquad \mbox{(AES)} \: .
\end{eqnarray}
The creator $b^\dagger_\beta$ refers to the core state.

Finally, an electron ($\alpha$) approaching the crystal surface can excite 
a core electron ($\beta$) into a state of the unoccupied bands. 
Above a threshold energy both, the de-excited primary electron and the 
excited core electron occupy valence states ($\gamma$, $\gamma'$) above 
the Fermi energy after the transition.
The appearance-potential spectroscopy (APS, $s=-2$) monitors the intensity 
of this transition as a function of polarization, energy and momentum of the 
incoming electron beam by detecting the emitted x-rays or Auger electrons 
of the subsequent core-hole decay. 
The perturbation is:
\begin{eqnarray}
  V_{-2} = \sum_{\alpha \beta \gamma' \gamma}
  M_{\gamma'\gamma\alpha\beta}^\ast \, 
  c^\dagger_{\gamma'} c^\dagger_\gamma b_\beta a_\alpha  
  + \mbox{h.c.} \qquad \mbox{(APS)} \: .
\end{eqnarray}
Again, $V_{-2}=V_2$. Since AES and APS are induced by the Coulomb 
interaction, the matrix element reads:
\begin{equation}
  M_{\alpha\beta\gamma'\gamma} = 
  {}^{(1)} \! \langle \alpha | \,
  {}^{(2)} \! \langle \beta  | \:
  \frac{1}{|{\bf r}_1-{\bf r}_2|} \:
  | \gamma' \rangle^{(1)}  \,
  | \gamma  \rangle^{(2)} \qquad (s=\pm 2) \: .
\end{equation}

To calculate the cross sections for the different spectroscopies, 
time-de\-pen\-dent perturbation theory with respect to the perturbation
$V_s$ is applied.
Using Fermi's golden rule, the transition probability per unit time 
for a transition from the initial state $|E_i\rangle$ to the final state
$|E_f\rangle$ is given by:
\begin{equation}
  w_s = 2 \pi \, | \langle E_f | V_s | E_i \rangle |^2
        \, \delta(E_f - E_i - \omega_s) \: .
\end{equation}
$|E_i\rangle$ and $|E_f\rangle$ are eigenstates of the grand-canonical
Hamiltonian $H-\mu N$ with eigenenergies $E_i$ and $E_f$. 
$\mu$ is the chemical potential and $N$ the particle-number operator.
Furthermore, $\omega_s = \pm \omega$ for $s=\pm 1$ where
$\omega$ is the photon frequency.
($\omega_s = 0$ for $s=\pm 2$). 

We proceed by applying the so-called sudden approximation which reads:
\begin{eqnarray}
  | E_f \rangle &= a^\dagger_\alpha | E_n \rangle \: , \qquad
  E_f &= \epsilon_\alpha + E_n
  \qquad (s=+1,+2) \: ,
\nonumber \\
  | E_i \rangle &= a^\dagger_\alpha | E_n \rangle \: , \qquad
  E_i &= \epsilon_\alpha + E_n
  \qquad (s=-1,-2) \: .
\end{eqnarray}
At this point we have neglected the interaction of the high-energy electron
with the rest system. 
The latter is left in the $n$-th excited state of $H$ with eigenenergy $E_n$.
$\epsilon_\alpha$ is the one-particle energy of the high-energy scattering
state.
The electron and the rest system propagate independently in time but 
consistent with energy conservation.
Generally, the sudden approximation is believed to hold well if
$\epsilon_\alpha$ is not too small.
We can furthermore assume that 
\begin{equation}
  a_\alpha | E_i \rangle \simeq 0 \qquad (s=+1,+2) \: , \qquad \quad
  a_\alpha | E_f \rangle \simeq 0 \qquad (s=-1,-2) \: .
\end{equation}
Hence, 
\begin{eqnarray}
  \langle E_f | V_s | E_i \rangle =
  \langle E_n | \: [a_\alpha , V_s]_- \: | E_i \rangle =
  \langle E_n | T_\alpha^{(s)} | E_i \rangle
  \qquad && (s=+1,+2) \: ,
\nonumber \\
  \langle E_f | V_s | E_i \rangle =
  \langle E_f | \: [V_s,a^\dagger_\alpha]_- \: | E_n \rangle =
  \langle E_n | {T_\alpha^{(s)}}^\dagger | E_i \rangle
  \qquad && (s=-1,-2) \: ,
\end{eqnarray}
where $[...,...]_-$ denotes the commutator and
\begin{eqnarray}
  T_\alpha^{(s)} \equiv [a_\alpha , V_s]_- = 
  \sum_{\gamma} M_{\alpha\gamma} \, a_\alpha^\dagger c_\gamma
  \qquad && (s = \pm 1) \: ,
\nonumber \\
  T_\alpha^{(s)} \equiv [a_\alpha , V_s]_- = 
  \sum_{\beta \gamma' \gamma}
  M_{\alpha\beta\gamma'\gamma} \, b^\dagger_\beta c_\gamma c_{\gamma'} 
  \qquad && (s = \pm 2) 
\label{eq:trop}
\end{eqnarray}
the transition operator.
The transition probability per unit time now reads:
\begin{eqnarray}
  w_s = 2 \pi \, | \langle E_n | T_\alpha^{(s)} | E_i \rangle |^2 \,
  \delta(E_n + \epsilon_\alpha - E_i - \omega_s) \qquad && (s=+1,+2) \: , 
\nonumber \\
  w_s = 2 \pi \, | \langle E_f | {T_\alpha^{(s)}}^\dagger | E_n \rangle |^2 \,
  \delta(E_f - E_n - \epsilon_\alpha - \omega_s) \qquad && (s=-1,-2) \: .
\end{eqnarray}

To get the intensity $I_s$ we have to average over the possible initial states.
Initially, the system is assumed to be in thermal equilibrium: 
At the temperature $T$ ($\beta=1/k_{\rm B}T$) the system is found in the
state $| E_i \rangle$ with the probability 
$W_i = Z^{-1} \exp(-\beta E_i)$.
Here $Z = \sum_i \exp(-\beta E_i)$ is the grand canonical partition function.
We consider an experiment that determines all quantum numbers $\alpha$
necessary for a complete measurement ($s=+1,+2$) or preparation ($s=-1,-2$) 
of the state of the high-energy electron.
Consequently, all indices have to be summed over except for $\alpha$.
Eventually, this yields the intensity:
\begin{eqnarray}
  I_s(\alpha) = \frac{2 \pi}{Z} \sum_{i,n} e^{-\beta E_i}
  | \langle E_n | T_\alpha^{(s)} | E_i \rangle |^2 \,
  \delta(E_n + \epsilon_\alpha - E_i - \omega_s) \quad && (s=+1,+2) \: , 
\nonumber \\
  I_s(\alpha) = \frac{2 \pi}{Z} \sum_{f,n} e^{-\beta E_n} 
  | \langle E_f | {T_\alpha^{(s)}}^\dagger | E_n \rangle |^2 \,
  \delta(E_f - E_n - \epsilon_\alpha - \omega_s) \quad && (s=-1,-2) \: .
\nonumber \\
\label{eq:intdirect}
\end{eqnarray}

These expressions can be written in a more compact form. 
For this purpose we define the Green function \cite{AGD64} as:
\begin{equation}
  G_{\alpha\alpha'}(E) = \frac{1}{Z} \sum_{m,n} 
  e^{-\beta E_n} (e^{\beta E}+1) 
  \frac{
  \langle E_n | {T_{\alpha'}^{(s)}}^\dagger | E_m \rangle
  \langle E_m | T_\alpha^{(s)} | E_n \rangle
  }
  {E-(E_n-E_m)} \: ,
\label{eq:green}
\end{equation}
which can also be written in the form $G_{\alpha\alpha'}(E)= \langle \langle 
T_\alpha^{(s)} ; {T_{\alpha'}^{(s)}}^\dagger \rangle \rangle $ to show the 
dependence on the transition operator.
With the help of the Dirac identity $1/(E+i0^+)={\cal P}(1/E) - i\pi\delta(E)$,
\begin{eqnarray}
  I_s(\alpha) = -2 \: \frac{1}{e^{\beta E}+1} \: \mbox{Im} \:
  G_{\alpha\alpha}(E+i0^+)
  \Big|_{E=\epsilon_\alpha-\omega_s} \quad && (s=+1,+2) \: , 
\nonumber \\
  I_s(\alpha) = -2 \: \frac{e^{\beta E}}{e^{\beta E}+1} \: \mbox{Im} \: 
  G_{\alpha\alpha}(E+i0^+)
  \Big|_{E=\epsilon_\alpha-\omega_s} \quad && (s=-1,-2) \: .
\label{eq:intgreen}
\end{eqnarray}
One recognizes the Fermi function $f(E)=1/(\exp(\beta E)+1)$ 
and the fact that both, the direct
($s>0$) and the inverse spectroscopies ($s<0$) are described by the same
(Green) function.
In fact, the following relations hold:
\begin{equation}
  I_{+1}(E) = e^{-\beta E} I_{-1}(E) \: , \qquad
  I_{+2}(E) = e^{-\beta E} I_{-2}(E) \: .
\end{equation}
Eqs.\ (\ref{eq:green}) and (\ref{eq:intgreen}) show that the intensity is 
given by a weighted sum of $\delta$ peaks at the excitation energies 
$E=E_n-E_m$. 
In the thermodynamic limit the energy spectrum will be continuous in 
general and thus the dependence $I_s(E)$ is smooth.
The weight factors $\langle E_n | {T_{\alpha'}^{(s)}}^\dagger | E_m \rangle
\langle E_m | T_\alpha^{(s)} | E_n \rangle$ distinguish between the
different spectroscopies.

The final equation (\ref{eq:intgreen}) is the goal of our considerations 
so far.
It relates the intensity to the Green function (\ref{eq:green}) which is
a central quantity of many-body theory. It
can be (approximately) calculated by standard diagrammatic methods such a 
perturbation theory with respect to the interaction strength or by methods 
involving infinite re-summations of diagrams \cite{AGD64}.
This is an essential advantage compared with a direct evaluation of 
Eq.\ (\ref{eq:intdirect}).
The latter seems to be impossible since one would have to compute explicitly
the eigenenergies $E_n$ and eigenstates $| E_n \rangle$ of a system of 
interacting electrons.

\section{APS for band ferro-magnets}
\label{sec:ferro}

In the following we will concentrate on the appearance-potential spectroscopy
to give an example how calculations can be performed in practice. 
Beforehand, however, some preparations are necessary.

\paragraph{Core-hole effects.}

A characteristic feature of APS is the formation of a core hole.
The energy of core level $\epsilon_c$ involved in the transition determines 
the shallow energy:
Energy conservation requires that the energy loss of the primary electron
must be equal to or larger than $E_{\rm F}-\epsilon_c$ where $E_{\rm F}$ 
is the Fermi energy.
Besides this static effect there are also dynamic core-hole effects 
originating from the scattering of valence electrons at the local core-hole
potential in the final state for APS \cite{PBNB93,PBB94}.
The dynamic core-hole effects are usually neglected assuming the Coulomb 
correlation between valence and core electrons to be small and not to 
affect the AP line shape significantly.
This is an approximation which is difficult to justify and which has to 
be checked in each case separately.

It leads, however, to a substantial simplification of the problem.
Similar to the sudden approximation discussed above, one can write:
\begin{eqnarray}
  | E_i \rangle = a^\dagger_\alpha b^\dagger_\beta | E_n \rangle' \: , \qquad
  E_i = \epsilon_\alpha + \epsilon_c + E'_n \: , \qquad
  b_\beta | E_f \rangle = 0
  \: .
\end{eqnarray}
With essentially the same steps as above one gets:
$I_{\rm APS} \propto (1-f(E)) \: \mbox{Im} G(E+i0^+)$ where now
$E=\epsilon_\alpha+\epsilon_c$ and where $G$ is a two-particle Green 
function of the type
$G = \langle \langle \: \sum M c_\gamma c_{\gamma'} \: ; 
\sum M^\ast c^\dagger_\gamma c^\dagger_{\gamma'} \rangle \rangle$. 
Note that the core-electron creators are eliminated 
(cf.\ Eq.\ (\ref{eq:trop})).
The only dynamic degrees of freedom left are those of the valence electrons
($\gamma$).

\paragraph{Hamiltonian.}

Characteristic for the valence electronic structure of the band-ferromagnetic 
$3d$ transition metals are the strongly correlated $3d$ bands around the Fermi 
energy which hybridize with essentially uncorrelated $4s$ and $4p$ bands.
The band structure derives from a set of localized one-particle basis states 
$| \gamma \rangle$ with $\gamma$ specified as $\gamma=(i,L,\sigma)$.
Here $| i L \sigma \rangle$ is taken to be a localized (atomic-like) orbital 
centered at the site $i$ of a lattice with cubic symmetry.
$\sigma=\uparrow,\downarrow$ is the spin index. 
$L$ is the orbital index running over the five $3d$ orbitals, the $4s$ and the
three $4p$ orbitals.
We also introduce an index $m$ which labels the different $d$ orbitals, namely
the three-fold degenerate $t_{\rm 2g}$ and the two-fold degenerate $e_{\rm g}$ 
orbitals.
Using these notations the Hamiltonian $H$ reads:
\begin{equation}
  H = \sum_{ii'LL'\sigma} t_{ii'}^{LL'} \,
  c_{iL\sigma}^\dagger
  c_{i'L'\sigma} 
  + \frac{1}{2} \sum_{i\sigma \sigma'}\sum_{m_1 \ldots m_4} \! 
   U_{m_1 m_2 m_4 m_3} \,
   c_{im_1\sigma}^\dagger
   c_{im_2\sigma'}^\dagger 
   c_{im_3\sigma'} 
   c_{im_4\sigma} \: .
\label{eq:ham}
\end{equation}
This is a multi-band Hubbard-type model including a strongly screened on-site 
Coulomb interaction among the $d$ electrons.
The hopping term $\propto t_{ii'}^{LL'} = \langle i L \sigma | \, 
H(U_{\ldots}=0) \, | i' L' \sigma \rangle$ describes the ``free'' 
(non-interacting) band structure which can be obtained by Fourier transformation
to $\bf k$ space $t_{ii'}^{LL'} \mapsto t^{LL'}({\bf k})$ and subsequent
diagonalization $t^{LL'}({\bf k}) \mapsto \epsilon_r({\bf k})$.

\paragraph{AP intensity.}

Having specified the one-particle basis and the Hamiltonian, we can write
down the final expression for the AP intensity with all relevant 
dependencies made explicit:
\begin{eqnarray}
  I_{\sigma_c\sigma_i} ({\bf k}_\|,E) & \propto &
  \mbox{Im} 
  \sum_{L_1L_2L_1'L_2'} 
  M_{L_1L_2}^{\sigma_c\sigma_i}({\bf k}_\|,E) \times
\nonumber \\ && \times
  \langle \langle 
    c_{iL_1\sigma_c} 
    c_{iL_2\sigma_i} ; 
    c_{iL_2'\sigma_i}^\dagger 
    c_{iL_1'\sigma_c}^\dagger 
  \rangle \rangle_E \;
  ( M_{L_1'L_2'}^{\sigma_c\sigma_i}({\bf k}_\|,E) )^\ast
  \: .
\label{eq:int}
\end{eqnarray}
The intensity depends on the quantum numbers $\beta$ of the core hole
formed, particularly on the spin $\sigma_c$ of the core state, and on 
the quantum numbers $\alpha$ of the incoming electron, its energy $E$,
its momentum parallel to the crystal surface ${\bf k}_\|$ and its spin
$\sigma_i$.

The AP line shape essentially results from intra-atomic transitions.
Consequently, those transition-matrix elements $M$ that lead to off-site 
contributions to the intensity are neglected in (\ref{eq:int}).
The ``raw spectrum'' as given by the imaginary part of the Green function
in Eq.\ (\ref{eq:int}) is therefore isotropic.
Any angular (${\bf k}_\|$) dependence is due to the angular dependence of
the high-energy scattering state in the matrix element.

The energy dependence of the intensity, i.~e.\ the actual AP line shape,
is mainly determined by the two-particle Green function and reflects
the energy-dependent probability for two-particle excitations.
On the contrary, for typical kinetic energies of the primary electron of 
the order of keV the change of the matrix elements due to the energy 
dependence of the high-energy scattering states is expected to be weak 
over a few eV.

\paragraph{Spin dependence.}

The Coulomb interaction that induces the transition conserves the electron
spin orientation.
So the spin orientations of the incoming electron $\sigma_i$ and of the
core electron $\sigma_c$ in the initial state determine the spin orientations
of the two additional valence electrons in the final state.
For an incoming electron with spin orientation $\sigma_i$ one can distinguish
between a ``singlet'' transition, i.~e.\ excitation of a core electron with 
$\sigma_c = -\sigma_i$, and a ``triplet'' transition with $\sigma_c = \sigma_i$.
It is important to note that the ratio between singlet and triplet transitions
is regulated by the symmetry behavior of the transition-matrix elements under 
exchange of the orbital indices.
Assume, for example, that the matrix element is symmetric: 
$M_{L_1L_2} = M_{L_2L_1}$. 
Then the transition operator for triplet transitions vanishes,
$T_{\sigma_c \sigma_i} = \sum_{L_1L_2} \, M_{L_1L_2} \, c_{iL_1\sigma_c} 
c_{iL_2\sigma_i} = 0$ for $\sigma_i = \sigma_c$,
since $c_{iL\sigma}^2=0$ (Pauli principle) and since
$c_{iL_1\sigma_c} c_{iL_2\sigma_i}$ is antisymmetric with respect
to $L_1 \leftrightarrow L_2$.
Thus, a symmetric or even a constant matrix element (as is sometimes assumed for 
simplicity) completely excludes the triplet transitions.

Consider a paramagnetic material for a moment. 
The argument above shows that
a symmetric $M_{L_1L_2}$ would imply $\sigma_c=-\sigma_i$, i.\ e.\ a fully 
polarized 
primary electron beam leads to a full polarization of the core hole.
Thus, any deviation from full core-hole polarization must be due to the 
antisymmetric part of the matrix element.

Below we will consider a ferromagnetic material and a situation where the 
spin state of the final core hole is not detected. 
Then, the intensities have to be summed incoherently:
$I_{\sigma_i} \equiv I_{\sigma_c\sigma_i} + I_{-\sigma_c\sigma_i}$. 
For a ferromagnetic material one expects the intensity to be still dependent 
on the spin orientation of the primary electrons:
$I_{\uparrow} \ne I_{\downarrow}$ (while $I_{\uparrow} = I_{\downarrow}$
above the Curie temperature).
Let us assume again symmetric behavior: $M_{L_1L_2} = M_{L_2L_1}$.
It has been argued above that this implies $I_{\sigma_c \sigma_i} = 0$ for 
$\sigma_i = \sigma_c$.
Furthermore, it is easy to see that 
$T_{\uparrow\downarrow} = - T_{\downarrow\uparrow}$ which implies
$I_{\downarrow \uparrow} = I_{\uparrow \downarrow}$.
Hence, $I_{\uparrow} = I_{\downarrow}$.
In conclusion, any spin asymmetry in the intensity is due to a non-vanishing
antisymmetric part of the matrix elements.
The AP intensity asymmetry is much more determined by the symmetry properties 
of the matrix elements with respect to their orbital indices as compared to 
their spin dependence.

\paragraph{Direct and indirect correlations.}

Any diagrammatic approach gives the two-particle Green function 
$\langle \langle c \, c \, ; c^\dagger c^\dagger \rangle \rangle$ 
as a (complicated) functional ${\cal F}$ of the one-particle Green function
$\langle \langle c \, ; c^\dagger \rangle \rangle$.
One can thus distinguish between direct and indirect correlations \cite{PBB94}.
As has been shown above, the one-particle Green function corresponds 
to the (inverse) photoemission spectrum.
Indirect correlations in APS are those which originate from the 
renormalization of the free one-particle spectrum by the interaction term 
in (\ref{eq:ham}).
The direct correlations, on the other hand, are represented by the concrete 
form of the functional ${\cal F}$.
Essentially, the direct correlations originate from the direct Coulomb
interaction of the two additional valence electrons in the final state. 
Since both electrons are created at the same site, they are affected by
the strong intra-atomic interaction. 
This may give rise to correlation-induced satellites in the AP spectrum 
as is demonstrated by the so-called Cini-Sawatzky theory \cite{Cin77,Saw77}.

When neglecting the direct correlations, the functional ${\cal F}$ becomes
a mere self-convolution of the one-particle Green functions. 
If furthermore the tran\-si\-tion-matrix elements are taken to be 
constant, the 
AP spectrum is simply given by the self-convolution of the unoccupied part 
of the one-particle density of states 
$\propto \mbox{Im} \langle \langle c \, ; c^\dagger \rangle \rangle$.
This is the so-called Lander model \cite{Lan53} which is frequently employed
for a rough interpretation of the spectra.

Within the framework of the self-convolution (Lander) model the two
final-state electrons propagate independently.
As a consequence one finds that only the direct term with $L_1=L_1'$ and 
$L_2=L_2'$ and the exchange term with $L_1=L_2'$ and $L_2=L_1'$ contribute 
to the sum over the orbital indices.
Generally, however, the Green function in Eq.\ (\ref{eq:int}) depends on 
{\em four} orbital indices. 
This implies that the usual characterization of the final state with two 
quantum numbers ($d$-$d$, $s$-$d$, etc.) is no longer valid if the direct
correlations are included.
The orbital character may change by electron scattering.

\section{Appearance-potential spectra of Nickel}
\label{sec:ni}

The significance of electron-correlation effects in APS shall be elucidated
in a more detailed discussion below.
For this purpose we concentrate on ferromagnetic Nickel as a prototypical 
$3d$ band-ferromagnet and compare the results of the theoretical approach
with experimental data.

\paragraph{Experiments.}

Experimental results are available for a Ni(110) single-crystal surface with 
in-plane magnetization \cite{Vonbank,PWN+00}.
In the setup a spin-polarized electron beam is used for excitation which
is emitted from a GaAs source. 
To correct for the incomplete polarization of the electrons ($P \approx 30\%$),
all data are rescaled to a 100\% hypothetical beam polarization.
The spin effect is maximized by alignment of the electron polarization 
and the sample magnetization vector.
The core-hole decay is detected via soft-X-ray emission (SXAPS). 
To separate the signal from the otherwise overwhelming background, modulation
of the sample potential by a peak-to-peak voltage of $2V$ together with lock-in 
technique is employed.
Details of the experimental setup can be found in Refs.\ \cite{Vonbank,EVDN93}.

\begin{figure}[t]
\begin{center}
\includegraphics[width=.62\textwidth]{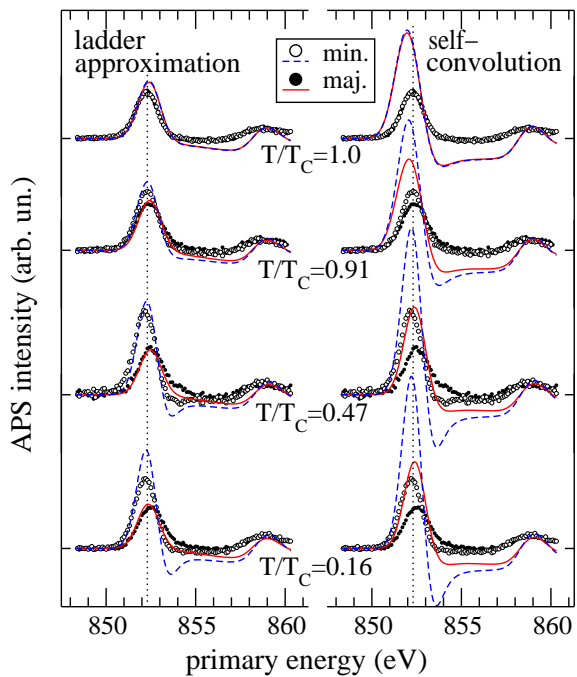}
\end{center}
\caption{
(from Ref.\ \cite{PWN+00})
Spin-resolved $L_{\rm III}$ Ni AP spectrum for 
different reduced temperatures $T/T_{\rm C}$.
{\em Data points:} measured differential intensity $dI/dE$ as a 
function of the primary energy. 
For better comparison with theory the same data are shown twice 
(left and right panel).
{\em Lines, left:} theory with direct and indirect correlations included 
(ladder approximation).
{\em Lines, right:} indirect correlations included only (self-convolution)
}
\label{fig:aps}
\end{figure}

Fig.\ \ref{fig:aps} shows the measured differential AP intensity as a 
function of the energy of primary electrons with polarization parallel 
(minority, $\downarrow$) or antiparallel (majority, $\uparrow$) to the 
target magnetization.
The displayed energy range covers the emission from the $L_{\rm III}$ 
transition ($2p_{3/2}$ core state). 
The $L_{\rm II}$ ($2p_{1/2}$) emission would be seen at higher energies 
shifted by the $2p$ spin-orbit splitting of 17.2~eV.

For $T/T_{\rm C}=0.16$ ($T_{\rm C} \approx 630$~K) the system is close to 
ferromagnetic saturation.
The AP spectrum shows a strongly spin-asymmetric intensity ratio as well 
as a spin splitting of the main peak at $E=852.3$~eV (indicated by the 
dotted line).
Since Ni is a strong ferromagnet, there are only few unoccupied $d$ states 
available in the majority spin channel, and thus $I_\uparrow < I_\downarrow$ 
holds for the (non-differential) intensities. 
This is the dominant spin effect. 
The intensity asymmetry in the main peak gradually diminishes with increasing 
$T$ and vanishes at $T_{\rm C}$.

The main peak is related to the high DOS at the Fermi energy.
Within an independent-electron picture, one can thus characterize the main 
peak as originating from transitions with two final-state electrons of $d$-$d$ 
character mainly. 
This is corroborated by calculations based on DFT-LDA \cite{EVDN93,EP97}.
Additional small $s$-$d$ contributions are present in the secondary peak at 
$E \approx 859$~eV as has been concluded from the analysis of the 
transition-matrix elements.
The secondary peak has been identified as resulting from a DOS discontinuity 
deriving from the $L_7$ critical point in the Brillouin zone \cite{EVDN93}.
No temperature dependence and spin asymmetry is detectable here.

\paragraph{Hamiltonian.}

To study the significance of electron correlations, we consider a nine-band 
Hubbard-type model $H=H_0+H_1$ including correlated $3d$ and uncorrelated 
$4s$/$4p$ orbitals as given by Eq.\ (\ref{eq:ham}).
The first (``free'') term $H_0$ is obtained from a Slater-Koster fit to the 
paramagnetic LDA band structure for Ni \cite{WPN00}.
Opposed to PES/IPE, this comparatively simple tight-binding 
parameterization appears to be sufficient in the case of APS since the 
two-particle spectrum does not crucially depend on the details of the 
one-particle DOS.

The on-site interaction among the $3d$ electrons is described by the second
term $H_1$.  
Exploiting atomic symmetries the Coulomb-interaction parameters which depend 
on the four orbital indices $m_1,\ldots,m_4$ can essentially be expressed in 
terms of two independent parameters $U$ and $J$.
The numerical values for the direct and exchange interaction, $U=2.47$~eV and 
$J=0.5$~eV, are taken from Ref.\ \cite{WPN00} where they have been fitted to 
the ground-state magnetic properties of Ni.
Residual interactions involving de-localized $s$ and $p$ states are assumed to 
be sufficiently accounted for by the LDA.
Finally, a double-counting correction is applied to $H$ since partially the
interaction $H_1$ is already included in the LDA Hamiltonian $H_0$ (for details 
see Ref.\ \cite{WPN00}).

\paragraph{Green functions.}

The two-particle Green function in Eq.\ (\ref{eq:int}) is approximately 
calculated by using standard diagrammatic techniques.
Because of the low density of $3d$ holes in the case of Ni, it appears to be 
reasonable to employ the so-called ladder approximation \cite{Nol90} which 
extrapolates from the exact (Cini-Sawatzky) solution for the limit of the 
completely filled valence band \cite{Cin77,Saw77}. 
For finite hole densities the ladder approximation gives the two-particle as 
a functional $\cal F$ of the one-particle Green function (direct correlations). 

The one-particle Green function, which describes the indirect correlations, is 
calculated self-consistently within second-order perturbation theory (SOPT) 
around the Hartree-Fock solution \cite{WPN00}. 
For a moderate $U$ and a low hole density, a perturbational approach can be 
justified \cite{SAS92}. 
A re-summation of higher-order diagrams is important to describe bound states 
(``Ni 6 eV satellite'') \cite{Lie81} which, however, are relevant for AES only.
Since spin-wave excitations are neglected in the approach, the calculated 
Curie temperature $T_{\rm C}=1655$~K is about a factor 2.6 too high. 
Using reduced temperatures $T/T_{\rm C}$, however, the temperature trend of 
the magnetization is well reproduced \cite{WPN00}.

\paragraph{Matrix elements.}

The transition-matrix elements in Eq.\ (\ref{eq:int}),
\begin{equation}
  M_{L_1L_2}^{\sigma_c\sigma_i}({\bf k}_\|,E) =
  \langle \mbox{2p}, \sigma_c |           \:
  \langle {\bf k}_\| E \sigma_i |         \:
  r_{12}^{-1}                             \:
  | i L_1 \sigma_c \rangle                \:
  | i L_2 \sigma_i \rangle \: ,
\label{eq:tme}
\end{equation}
are calculated by assuming the transition to be intra-atomic as usual 
\cite{Ram91,HWMR88,EP97}. 
The different wave functions as well as the Coulomb operator $1/r_{12}$ are 
expanded into spherical harmonics, the angular integrations are performed 
analytically, and the numerical radial integrations are cut at the Wigner-Seitz 
radius. 

Surface effects enter the theory via the high-energy scattering state 
$|{\bf k}_\|E\sigma_i\rangle$.
It is calculated as a conventional LEED state with ${\bf k}_\|=0$ to describe 
the normally incident electron beam in the experimental setup.
The (paramagnetic) LDA potential for Ni is determined by a self-consistent 
tight-binding linear muffin-tin orbitals (LMTO) calculation \cite{AJ84}.
The $3d$, $4s$, and $4p$ valence orbitals $| i L \sigma \rangle$ are taken 
to be the muffin-tin orbitals.
The four-fold degenerate $2p_{3/2}$ core state is obtained from the LDA core 
potential by solving the radial Dirac equation numerically. 
Its (relativistically) large component is decomposed into a (coherent) sum of 
Pauli spinors $| \mbox{2p}, \sigma_c \rangle$ with 
$\sigma_c = \uparrow, \downarrow$.

\paragraph{Results.}

The solid lines in Fig.\ \ref{fig:aps} (left) show the spectra as 
calculated from Eq.\ (\ref{eq:int}) using the ladder approximation 
for the two-particle Green function.
To account for apparatus broadening, the results have been folded 
with a Gaussian of width $\sigma = 0.6$~eV (see Ref.\ \cite{EP97}). 
The calculated data are shifted by 852.3~eV such that onset of the 
un-broadened spectrum for $T/T_{\rm C}=0.16$ coincides with the maximum 
of $L_{\rm III}$ emission in the experiment (dotted line).
Fig.\ \ref{fig:aps} and also a more detailed inspection show that the 
secondary peak at $E \approx 859$~eV is not affected by correlations 
at all.
This is consistent with observed temperature independence of the peak 
and with the fact that the DOS has mainly $s$-$p$ character at the 
discontinuity deriving from the $L_7$ critical point.
The maximum of the secondary peak is therefore used as a reference to 
normalize the measured spectra for each temperature.

What are the signatures of electron correlations?
The indirect correlations manifest themselves as a renormalization of 
the one-particle DOS.
Here, they are responsible in first place for the correct temperature 
dependence of the intensity asymmetry of the main peak in the AP spectrum.
Details are discussed in Refs.\ \cite{WPN00,PWN+00}.
The the direct interaction between the two additional final-state 
electrons and thus the direct correlations are much more important 
for APS. 
To estimate this effect, Fig.\ \ref{fig:aps} (right) also displays 
the results of the self-convolution model for comparison 
(still including matrix elements as well as the fully interacting 
one-particle DOS).

Looking at the results of the ladder approximation, the overall 
agreement with the measurements is rather satisfying.
Except for the lowest temperature the intensity, the spin splitting 
and the spin asymmetry of the main peak are well reproduced and, 
consistent with the experiment, a negligibly small intensity asymmetry 
and spin splitting is predicted for the secondary peak.
However, switching off the direct correlations (Fig.\ \ref{fig:aps}, 
right), results in a strong overestimation of the main peak structure. 

A plausible qualitative explanation of this pronounced correlation 
effect can given within the Cini-Sawatzky theory:
For low hole density the main effect of the direct correlations is 
known to transfer spectral weight to lower energies inaccessible to 
APS. 
This weight shows up again in the (complementary) Auger spectrum 
(recall that APS and AES are described by the same Green function).
Hypothetically, for $U\mapsto \infty$ all weight would be taken by 
a satellite split off at the lower boundary of the Auger spectrum 
\cite{Cin77,Saw77}.
A considerable weight transfer is in fact seen in AES \cite{Ram91,BFHLS83}.

\begin{figure}[t]
\begin{center}
\includegraphics[width=.48\textwidth]{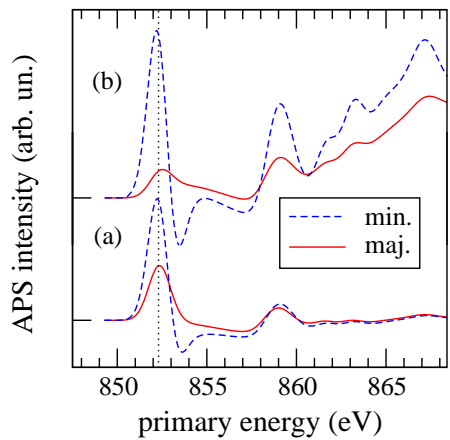}
\end{center}
\caption{
(from Ref.\ \cite{PWN+00})
Ni AP spectrum for $T=0$.
(a) full theory. 
(b) as (a) but matrix elements taken to be constant (see text)
}
\label{fig:tme}
\end{figure}

Fig.\ \ref{fig:tme} shows the effect of the transition-matrix elements.
Their importance for a quantitative understanding of spin-resolved APS 
can be demonstrated by setting 
$M_{L_1L_2}^{\sigma_c\sigma_i}({\bf k}_\|,E) = \pm 1 = \mbox{const}$ 
for $L_1 \ge L_2$ or $L_1 < L_2$, respectively, (see below) and comparing with 
the results of the full theory.
Their energy dependence (via the energy dependence of 
$| {\bf k}_\| E \sigma_i \rangle$) is weak over a few eV at energies of the 
order of keV and cannot explain the difference between (a) and (b) in 
Fig.\ \ref{fig:tme}.
The main difference is rather a consequence of the fact that the radial $2p$ 
core wave function has a stronger overlap with the (more localized) $3d$ as 
compared to the (more de-localized) $4s/4p$ radial wave functions.
This implies a suppression of the $s$-$p$ contributions to the orbital sum 
in Eq.\ (\ref{eq:int}).
The features above $E=860$~eV originate from additional discontinuities of 
the $s$-$p$-like DOS (as for the peak at $E \approx 859$~eV).

For $T<T_{\rm C}$ the spin asymmetry of the spectrum is mainly due to the 
spin dependence of the Green function in Eq.\ (\ref{eq:int}). 
If the calculation of the matrix elements (\ref{eq:tme}) starts from the 
{\em spin-polarized} L(S)DA potential, an additional spin asymmetry is 
observed resulting from the spin dependence of the states in Eq.\ (\ref{eq:tme}). 
This, however, is small and has practically no influence on the results.

On the other hand, Fig.\ \ref{fig:tme} shows a strong suppression of the 
intensity asymmetry at high energies when taking matrix elements into account.
This effect is controlled by the symmetry of the matrix 
$M_{L_1L_2} \equiv M_{L_1L_2}^{\sigma_c\sigma_i}({\bf k}_\|,E)$.
In the antisymmetric case, $M_{L_1L_2} = - M_{L_2L_1}$ ($L_1 \ne L_2$), there
is a maximum spin asymmetry (Fig.\ \ref{fig:tme}) while, even for a 
ferromagnet, there is no spin asymmetry at all for the symmetric case, as
discussed in section \ref{sec:ferro}.
The results of the full calculation along Eq.\ (\ref{eq:tme}) are neither 
fully symmetric nor antisymmetric with respect to $L_1$, $L_2$.

\section{Conclusions}
\label{sec:con}

It has been shown that the basic theory of different electron spectroscopies,
PES, IPE, AES and APS, can be developed within a unified framework.
Starting from Fermi's golden rule and employing the sudden approximation, 
the intensity can be expressed in terms of a one- or two-particle Green 
function including electron-photon or Coulomb matrix elements, respectively.
Citing APS from the typical band ferromagnet Ni as an example, it could 
be demonstrated that a quantitative agreement with spin-resolved and
temperature-dependent measured spectra can be achieved when the basic
spectroscopic formula is taken seriously and is evaluated by modern techniques 
of solid-state theory.

In particular, it has been demonstrated that there are pronounced (direct)
correlation effects in the AP line shape of Ni.
While $s$-$p$ derived features at higher energies appear to be sensitive 
to the geometrical structure only, the main peak is strongly affected by 
the direct interaction between the two additional final-state electrons. 
Consistent with the Cini-Sawatzky model and consistent with the well-known 
Ni Auger spectrum, there is a considerable spectral-weight transfer to 
energies below the threshold.
For Co and Fe one can even expect stronger effects of $d$-$d$ correlations
on the AP line shape since the $d$-hole density is larger than in Ni.
Simple self-convolution models neglecting the direct correlations must be 
questioned seriously.

It has also been demonstrated that a mere computation of the (two-particle)
Green function is insufficient to describe spin-resolved APS. 
The spin asymmetry of the AP signal is found to be mainly determined by
the orbital-dependent transition-matrix elements and their transformation 
behavior under exchange of the orbital quantum numbers.

Considering the temperature dependence of the spectra and the magnetic order 
resulting from strong (indirect) correlations in addition, one can state that 
the AP line shape of a typical ferromagnetic $3d$ transition metal is determined 
by a rather complex interplay of different factors.

Despite the fact that a reasonable understanding of APS from Ni has been 
achieved, there is much work to be done in the future:
An open question concerns the importance of core-hole effects in APS, for
example.
For the present case there has been no need to consider scattering at the
core-hole potential in the final state. 
This may likely be different for systems with a smaller $3d$ occupancy.
Furthermore, one must recognize that even the determination of the 
valence-band Green function is a central problem of many-body theory. 
Recently much progress has been achieved to deal with single-band models
\cite{GKKR96}; for a quantitative interpretation of electron spectra from 
transition metals, however, one needs a theory that realistically includes 
orbital degeneracy and $sp$-$d$ hybridization from the very beginning.
Compared to perturbation theory or lowest-order re-summation of diagrams, 
as employed here, improvements are conceivable and necessary albeit not 
easily performed.
\vfill

\section*{Acknowlegdements}

I would like to thank 
J. Braun, M. Donath (Universit\"at M\"unster), 
W. Nolting, T. Wegner (Humboldt-Universit\"at zu Berlin)
and T. Schlath\"olter (Philips Hamburg)
for stimulating discussions.


\begin{thebibliography}{10}
\addcontentsline{toc}{section}{References}

\bibitem{Pen74}
J.~B. Pendry: \emph{Low Energy Electron Diffraction}
  (Academic, London 1974)

\bibitem{PH74}
R.~L. Park, J.~E. Houston:
 J. Vac. Sci. Technol. {\bf 11}, 1 (1974)

\bibitem{FFW78}
\emph{Photoemission and the Electronic Properties of Surfaces}, ed. by
  B.~Feuerbacher, B.~Fitton, R.~F. Willis (Wiley, New York 1978)

\bibitem{Fug81}
J.~C. Fuggle:
  \emph{Electron Spectroscopy: Theory, Techniques and Applications}, vol. 4
  (Academic, London 1981) p. 85

\bibitem{Ram91}
D.~E. Ramaker:
 Crit. Rev. Solid State Mater. {\bf 17}, 211 (1991)

\bibitem{Don94}
M.~Donath:
 Surf. Sci. Rep. {\bf 20}, 251 (1994)

\bibitem{Pen76}
J.~B. Pendry:
 Surf. Sci. {\bf 57}, 679 (1976)

\bibitem{Bor85}
G.~Borstel:
 Appl. Phys. A {\bf 38}, 193 (1985)

\bibitem{Bra96}
J.~Braun:
 Rep. Prog. Phys. {\bf 59}, 1267 (1996)

\bibitem{HK64}
P.~Hohenberg, W.~Kohn:
 Phys. Rev. {\bf 136}, 864 (1964)

\bibitem{KS65}
W.~Kohn, L.~J. Sham:
 Phys. Rev. {\bf 140}, 1133 (1965)

\bibitem{PLNB97}
M.~Potthoff, J.~Lachnitt, W.~Nolting, J.~Braun:
 phys. stat. sol. (b) {\bf 203}, 441 (1997)

\bibitem{MPN+99}
C.~Meyer, M.~Potthoff, W.~Nolting, G.~Borstel, J.~Braun:
 phys. stat. sol. (b) {\bf 216}, 1023 (1999)

\bibitem{Lan53}
J.~J. Lander:
 Phys. Rev. {\bf 91}, 1382 (1953)

\bibitem{HWMR88}
G.~H{\"o}rmandinger, P.~Weinberger, P.~Marksteiner, J.~Redinger:
 Phys. Rev. B {\bf 38}, 1040 (1988)

\bibitem{EP97}
H.~Ebert, V.~Popescu:
 Phys. Rev. B {\bf 56}, 12884 (1997)

\bibitem{Kir84}
J.~Kirschner:
 Solid State Commun. {\bf 49}, 39 (1984)

\bibitem{EVDN93}
K.~Ertl, M.~Vonbank, V.~Dose, J.~Noffke:
 Solid State Commun. {\bf 88}, 557 (1993)

\bibitem{Cin77}
M.~Cini:
 Solid State Commun. {\bf 24}, 681 (1977)

\bibitem{Saw77}
G.~A. Sawatzky:
 Phys. Rev. Lett. {\bf 39}, 504 (1977)

\bibitem{AGD64}
A.~A. Abrikosow, L.~P. Gorkov, I.~E. Dzyaloshinski:
 \emph{Methods of Quantum Field Theory in Statistical Physics}
 (Prentice-Hall, New Jersey 1964).

\bibitem{PBNB93}
M.~Potthoff, J.~Braun, W.~Nolting, G.~Borstel:
 J. Phys.: Condens. Matter {\bf 5}, 6879 (1993)

\bibitem{PBB94}
M.~Potthoff, J.~Braun, G.~Borstel:
 Z. Phys. B {\bf 95}, 207 (1994)

\bibitem{Vonbank}
M. Vonbank:
  Spinaufgel{\"o}ste Appearance Potential Spektroskopie an
  3d-\"Uber\-gangs\-metallen. PhD Thesis, TU Wien (1992)

\bibitem{PWN+00}
M.~Potthoff, T.~Wegner, W.~Nolting, T.~Schlath\"olter, M.~Vonbank, K.~Ertl,
  J.~Braun, M.~Donath (to be published)

\bibitem{WPN00}
T.~Wegner, M.~Potthoff, W.~Nolting:
 Phys. Rev. B {\bf 61}, 1386 (2000)

\bibitem{Nol90}
W.~Nolting:
 Z. Phys. B {\bf 80}, 73 (1990)

\bibitem{SAS92}
M.~M. Steiner, R.~C. Albers, L.~J. Sham:
 Phys. Rev. B {\bf 45}, 13272 (1992)

\bibitem{Lie81}
A.~Liebsch:
 Phys. Rev. B {\bf 23}, 5203 (1981)

\bibitem{AJ84}
O.K. Andersen, O.~Jepsen:
 Phys. Rev. Lett. {\bf 53}, 2571 (1984)

\bibitem{BFHLS83}
P.~A. Bennett, J.~C. Fuggle, F.~U. Hillebrecht, A.~Lenselink, G.~A.
  Sawatzky:
 Phys. Rev. B {\bf 27}, 2194 (1983)

\bibitem{GKKR96}
A.~Georges, G.~Kotliar, W.~Krauth, M.~J. Rozenberg:
 Rev. Mod. Phys. {\bf 68}, 13 (1996)

\end{thebibliography}
\end{document}